\documentclass{iaus}
\usepackage{graphicx}

\title[Chemical spots on the surface of HgMn stars]
{Searching for a link between the presence of chemical spots on the surface
of HgMn stars and their weak magnetic fields}

\author[Savanov, Hubrig, Gonz\'alez, Sch\"oller]   
{I.S.~Savanov$^1$, 
S.~Hubrig$^2$, J.F.~Gonz\'alez$^3$ \and M.~Sch\"oller$^4$}

\affiliation{$^1$Institute of Astronomy, Russian Academy of Science,
Moscow, Russia \break email: isavanov@rambler.ru;
$^2$ESO, Santiago, Chile;
$^3$Complejo Astronomico El Leoncito, San~Juan, Argentina;
$^4$ESO, Garching, Germany
}
\pubyear{2009}
\volume{xxx}  
\pagerange{100--104}
\date{?? and in revised form ??}
\jname{Proceedings Title IAU Symposium}
\editors{A.C. Editor, B.D. Editor \& C.E. Editor, eds.}
\begin{document}

\maketitle

\begin{abstract}
We present the results of mapping the HgMn star AR Aur using the Doppler
Imaging technique for several elements and discuss the obtained
distributions in the framework of a magnetic field topology.
\end{abstract}

\firstsection 
\section{Introduction}

Late B-type stars with HgMn peculiarity are characterized by low
rotational velocities and weak or non-detectable magnetic
fields. The most distinctive features of their atmospheres are
the extreme overabundance of Hg and Mn. More than 2/3 of HgMn
stars belong to spectroscopic binaries. The presence of an
inhomogeneous distribution of some elements over the surface of
HgMn stars was discussed for the first time by \cite{Hubrig95}.
The first definitively identified spectrum variability
has been reported for the binary star $\alpha$ And by \cite{Wahlgren01}
and \cite{Adelman02}.

\section{Doppler imaging}\label{sec:Doppler imaging}

We present the first results of our preliminary
Doppler Imaging (DI) modeling of abundance
distributions for four elements on the surface
of the HgMn star AR~Aur (see Fig.~1).
The analysis of the spectra is made with the DI inversion
code iAbu, which reconstructs the stellar surface abundance
inhomogeneities from the series of spectral line profiles using the
Tikhonov regularization algorithm.

Previously \cite{Hubrig06} concluded that the employment of a
partially fractured equatorial ring presents a realistic
distribution of Y on the stellar surface. Our new map based on the
inversion of the Y~II $\lambda$4900\,\AA{} line is in good agreement with
this conclusion. The Y is overabundant and strongly concentrated in
the equatorial ring, where the average Y abundance reaches
log$\varepsilon$(Y) = 6.0 (in the scale log$\varepsilon$(H) =
12.0). One large fraction of the ring is missing exactly on
the surface area which is permanently facing the secondary.
An additional polar detail can be seen at longitudes
180$^{\circ}$--270$^{\circ}$.
The surface distribution of strontium was studied with the
ion line Sr~II $\lambda$4215\,\AA{}. The Sr surface distribution
in AR~Aur is not completely similar to the map reconstructed for
Y~II, although it also presents significant equatorial
details of Sr overabundance in the range of longitudes from
70$^{\circ}$ to 250$^{\circ}$.

In the case of the Hg~II $\lambda$3984\,\AA{} line we normalized our
spectra relative to the pseudo-continuum of the H$\varepsilon$ wing,
but accounted for blending by the hydrogen line in the
calculations of the synthetic spectra by adding corresponding
opacity sources. The line list used for the DI consists of
isotopic and hyperfine structure components of the Hg~II
$\lambda$3984\,\AA{} line and a Y~II $\lambda$3982.59\,\AA{} blending
line. 
We parameterized the
mercury isotope composition using the isotopic-fractionation
model of \cite{White76}. Our observational data are
insufficient for a simultaneous determination of both the Hg
abundance and the surface distribution of the q-parameter, hence we
accept q=0 for the terrestrial mixture.
The mercury overabundance is strongly
concentrated in several equatorial details on the surface of
the star. Further we find polar appendages at longitudes
0$^{\circ}$--90$^{\circ}$

\begin{figure}
\centering
\resizebox{4.4cm}{!}{\includegraphics{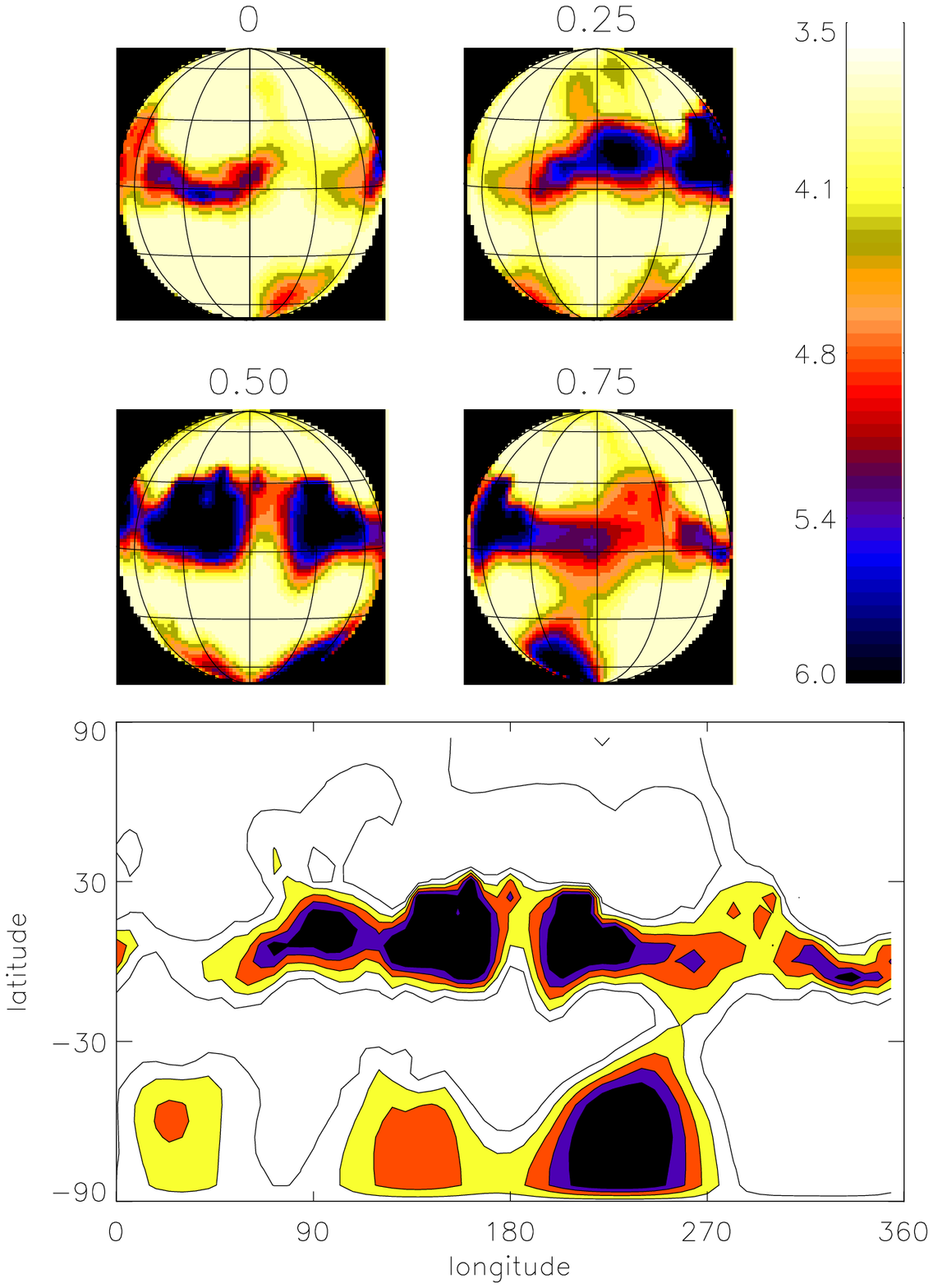} }
\resizebox{4.4cm}{!}{\includegraphics{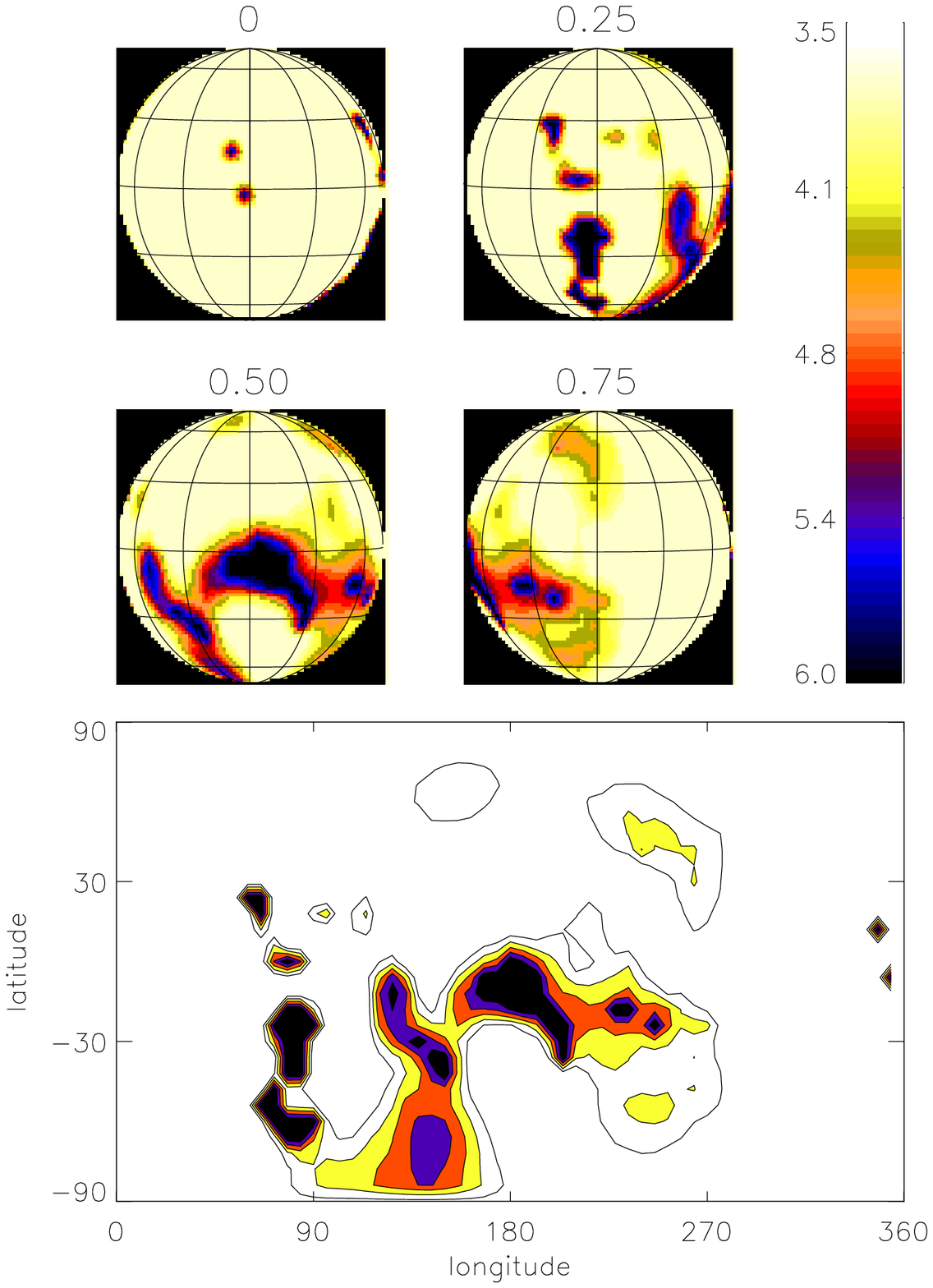} }
\resizebox{4.4cm}{!}{\includegraphics{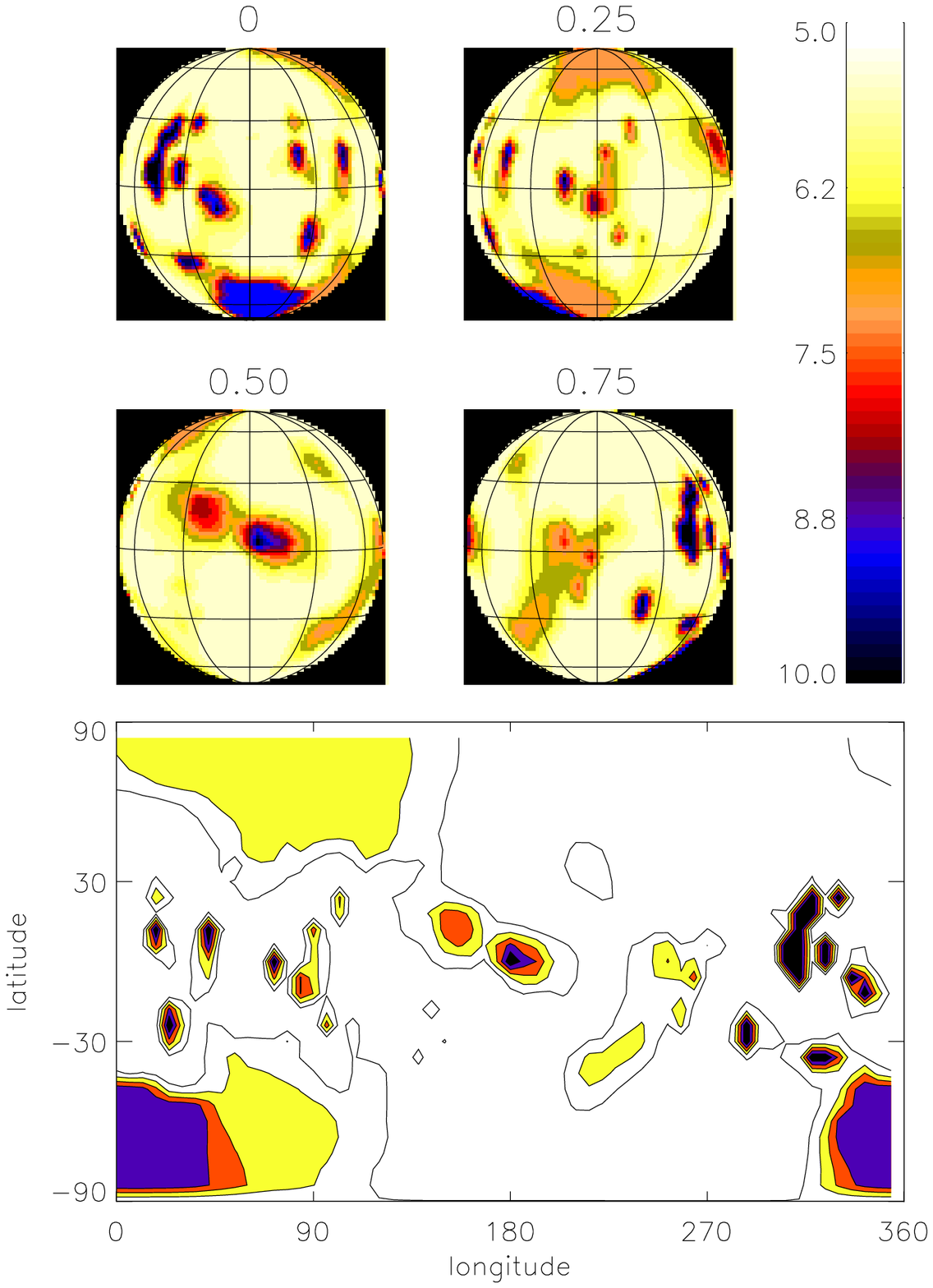} }
\caption[]{Results of the DI reconstruction of the Y,
Sr and Hg surface
distributions (from left to right).
The rectangular plots (lower panels) show a
pseudo-Mercator projection of the surface maps. Spherical DI maps 
are presented at four equidistant rotational phases (upper panels). }
\end{figure}

For the first time we reconstructed in a HgMn star the
manganese distribution on the stellar surface. 
The Mn ion map is based on modeling of the Mn~II $\lambda$4292\,\AA{} line.
This map is not presented in Fig.~1 due to lack of space.
We demonstrated that the Mn surface distribution shows similarities
with those of Y and Sr, namely both equatorial and polar features.


\section{Conclusions}\label{sec:concl}

What is the explanation for the discovered inhomogeneities?
Taking into account that more than 2/3
of the HgMn stars are known to belong to spectroscopic binaries,
a scenario how a magnetic field can be built up in binary systems has been presented some time ago by
\cite{Hubrig98} who suggested that a tidal torque varying with depth and
latitude in a star induces differential rotation. Differential
rotation in a radiative star can, however, be prone to the magneto-rotational
instability (e.g., \cite{Arlt03}). Magnetohydrodynamical
simulations revealed a distinct structure for the
magnetic field topology similar to the latitudinal fractured rings
observed on the surface of $\alpha$~And and AR~Aur.


\end{document}